\documentclass[11pt,a4paper]{article}
\usepackage{graphicx}  
\usepackage{amsmath,amssymb}

\begin{document}

\title{Quantum and classical localisation and the Manhattan lattice}
\author{E.~J.~Beamond$^1$, A. L. Owczarek$^2$ and J. Cardy$^{1,3}$\\
$^1$Theoretical Physics, University of Oxford,\\ 1 Keble
Road, Oxford  OX1 3NP, United Kingdom\\ 
$^2$Department of Mathematics and Statistics,\\ University of Melbourne,
Vic, 3010, Australia.\\  
$^3$All Souls College,\\ Oxford, United Kingdom}
\date{\today}
\maketitle

\begin{abstract}
We consider a network model, embedded on the Manhattan lattice, of a
quantum localisation problem belonging to symmetry class C. This
arises in the context of quasiparticle dynamics in disordered
spin-singlet superconductors which are invariant under spin rotations
but not under time reversal. A mapping exists between problems
belonging to this symmetry class and certain classical random walks
which are self-avoiding and have attractive interactions; we exploit
this equivalence, using a study of the classical random walks to gain
information about the corresponding quantum problem. In a
field-theoretic approach, we show that the interactions may flow to
one of two possible strong coupling regimes separated by a transition:
however, using Monte Carlo simulations we show that the walks are in
fact always compact two-dimensional objects with a well-defined
one-dimensional surface, indicating that the corresponding quantum
system is localised.

\vspace{1cm} 

\noindent{\bf Short Title:} Localisation and the Manhattan lattice


\end{abstract} 

\vfill

\newpage

\section{Introduction}
\label{intro}
Disordered electronic systems, and the associated metal-insulator
transitions, have generated much interest in recent years. In this
paper, we consider a two-dimensional system of non-interacting
particles which is invariant under spin rotations but not under time
reversal. According to the classification set out by Altland and
Zirnbauer \cite{alt-zirn} such a system belongs to symmetry class
C. This class is one of several which are distinct from the three
Wigner-Dyson classes and have realisations as Bogoliubov-de Gennes
Hamiltonians for quasiparticles in disordered superconductors. In
particular, class C arises in spin-singlet superconductors with broken
time-reversal symmetry for orbital motion but negligible Zeeman
splitting \cite{alt-zirn}.

\begin{figure}
\begin{center}
\includegraphics[width=14cm]{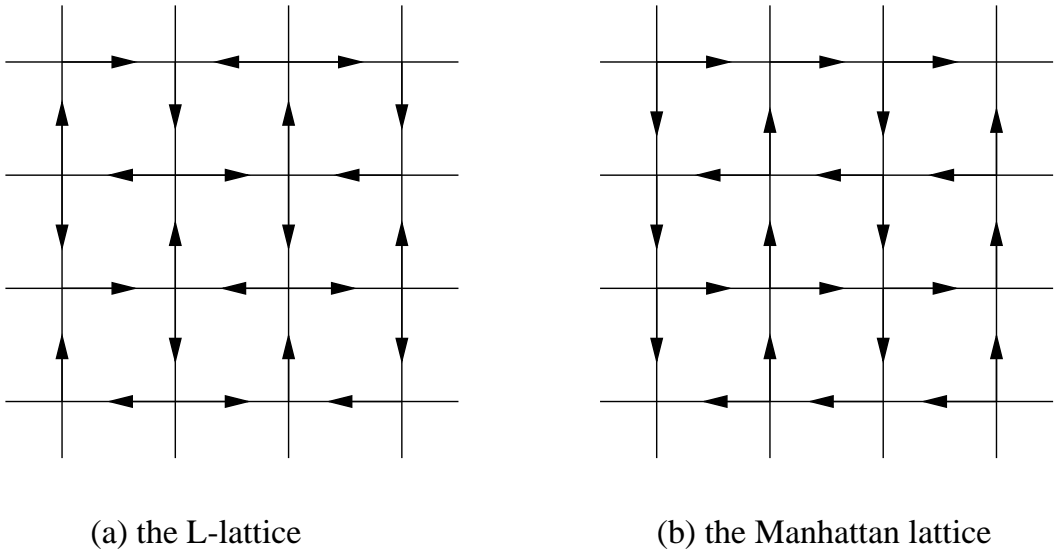}
\caption[The L and Manhattan lattices]{The L and Manhattan lattices}
\label{lattices}
\end{center}
\end{figure}

For two-dimensional systems with broken time reversal symmetry,
whether from the Wigner-Dyson unitary class or from class C, there
exists the possibility of a delocalisation transition of the quantum
Hall type. Recent studies of network models for systems belonging to
class C \cite{kag-prb, kag-prl, grl, smf, cardy-sqhe, bcc} have
focussed on this transition, yielding several interesting
results. Notably, one such network model describes the spin quantum
Hall effect, with two insulating phases separated by a delocalisation
transition analogous to the quantum Hall plateau transition. The two
phases exhibit quantised Hall conductance differing by an integer. It
has been shown \cite{grl, bcc} that this quantum problem may be mapped
to that of the hulls of classical percolation clusters in two
dimensions, a well-understood problem. The particular network model
used in these studies was based on the L-lattice (see
Fig.~\ref{lattices}a), and is a natural generalisation of the
Chalker-Coddington model\cite{chalk-codd} which was designed to study
the ordinary integer quantum Hall transition.

The L-lattice is a square lattice consisting of directed edges and
nodes, such that every path following the directed edges must turn left
or right at each node. It possesses a particular symmetry under inversion
about a diagonal axis which reflects the symmetry of the corresponding
quantum hamiltonian about a particular energy. 

However, in \cite{bcc} it was shown that the mapping of network models
in class C to classical random walk models is more general, and holds
on any directed lattice (at least if each node is of degree 4.)  These
walks are in general \emph{trails}, in that they may not pass over a
given link more than once, but they also have attractive interactions
at each node.  In the mapping, localised and extended phases of the
quantum system map to regions of parameter space where the
corresponding classical walks either close after a finite number of
steps (almost surely) or they escape to infinity.  The localisation
behaviour exhibited by a given network model, and the nature of the
corresponding classical problem is expected to be strongly dependent
on the structure of the underlying lattice. Therefore, by studying
classical problems formulated on different lattices, we might uncover
different types of localisation behaviour, resulting in the
competition between the edge exclusion and nodal attraction.

The classical random walks in which we are interested, together with
other kinetic self-avoiding trails, correspond to the classical
scattering of light by random arrays of mirrors laid out on a square
lattice. Problems of this type have been studied extensively
\cite{gunn-ort, kong, cohen, wang}. These processes may also be 
viewed as history-dependent kinetic random walks. For example, in terms of
the mirror model, the walker lays down a mirror at random on visiting
a node; on return to that node, it finds the mirror already fixed
there. Therefore, models of this type may be classified as ``true"
self-avoiding walks. This class of walks has been studied in the limit
of weak scattering using field-theoretic renormalisation group methods
\cite{amit,obukhov,peliti}; we shall make use of such techniques in
section \ref{analytic} of this paper to study such walks.
We note that, as for Anderson localisation, the
critical dimension for these processes is two.

As mentioned above, the network model describing the spin quantum Hall
effect is formulated on the L lattice (Fig.~\ref{lattices}a).  In the
heuristic justification for studying this lattice in this context (and
for the original Chalker-Coddington model) it is seen as a distorted
version of a random lattice, whose plaquettes correspond to the
regions where locally either $E-V$ is $<0$ or $>0$, where $E$ is the
Fermi energy and $V$ is some slowly varying random potential. In the
semi-classical approximation, there are edge states propagating along
the curves which divide these regions. Along these, the quasi-particle
wave function suffers a constant spin (or phase) rotation. On
distorting the random geometry to the L-lattice, these become quenched
random link variables of the model.

In this paper, we consider a different lattice, the Manhattan lattice.
It shares with the L-lattice the property that at each node there are
2 incoming and 2 outgoing edges or links, so that the theorems of
\cite{bcc} apply.  
On this lattice, a walker may either continue straight on at a given
node with a given probability or make a turn. The allowed directions
for turning at each node alternate right and left, as shown in
Fig.~\ref{lattices}b. The overall lattice is therefore regular, with a
similar size unit cell to the L-lattice.
In the network model, at each node there is a real unitary $S$-matrix
which describes scattering from the incoming edges to the outgoing ones.
Just as for the models on the L-lattice, this is in principle also a
quenched random matrix, but for simplicity we shall take it to be
the same at all nodes. Thus the $S$-matrix elements for continuing
straight on at each node are taken to be $\cos\theta$, and those for
turning are $\pm\sin\theta$. 
When $\cos\theta=1$, the quasiparticles move ballistically along the
lattice axes, while for $\sin\theta=\pm1$ they move around the perimeter
of single plaquettes. The existence of a ballistic limit distinguishes
this lattice from the L-lattice. 
On the other hand, the Manhattan
lattice appears to enjoy no special symmetry, as does the L-lattice
about $\theta=\pi/4$. 

The general results of \cite{bcc} show that the single-particle
Green's function and the two-point conductance for a class C model on
this lattice are related to the properties of classical trails, or
random walks which visit each edge no more than once, and, on first
visiting a given node, turn L or R with probability $p\equiv
\sin^2\theta$, or continue straight with probability
$1-p=\cos^2\theta$.  For the L-lattice, the limits $p\rightarrow 0$
and $p\rightarrow 1$ show clearly two distinct behavioural phases,
characterised by different edge states; a delocalisation transition
occurs at $p=1/2$. For the model formulated on the Manhattan lattice,
however, we might expect to find different behaviour, since it is
clear that there are no corresponding edge states as $p\rightarrow
0,1$.  We note also that a kinetic self-avoiding trail on the
Manhattan lattice may cross itself with probability $p$, but this is
never permitted on the L lattice.

At $p=1$, the walks are simply loops defining the elementary
plaquettes of the lattice. In fact, it may be argued\cite{jtc} that
the trajectories are loops of finite extent at least for $p>\frac12$:
if we place a mirror at 45$^\circ$ at each of the vertices where the
path turns through $\pm90^\circ$, they may be thought of as forming
clusters of bonds on a square lattice at 45$^\circ$ to the
original. For $p>\frac12$ the mirrors percolate: that is, they form an
infinite cluster which crosses the entire lattice, with probability
one. The trajectories are confined to the voids in this cluster, which
are finite. For $0<p\leq\frac12$, this argument is inconclusive: the
trajectories will scatter from the finite clusters of mirrors, but
this may or may not lead to classical localisation.  At $p=0$ all
trajectories are straight lines; the states to which they correspond
are spatially extended.  Therefore either almost all trajectories are
localised for all $p>0$, or there is a transition at some finite value
$0<p_c\leq\frac12$ from extended to localised behaviour.

In this paper, we use two approaches to study the kinetic processes
described above. In section \ref{analytic}, we develop a
field-theoretic description of the weak-coupling regime of this
particular type of ``true" self-avoiding walk. We begin by estimating
the diffusion constant for the Manhattan lattice, and show that the
problem is in this regime as $p\to0$. We then use field-theoretic
renormalisation group methods to analyse the general class of such
problems: the RG trajectories always run away from the weak-coupling
fixed point, with possibility of two distinct strong-coupling
behaviours separated by a critical trajectory.  However, in section
\ref{simulations}, we use Monte-Carlo methods directly to simulate kinetic
self-avoiding trails on the Manhattan lattice. These simulations are
only practical for values of $p$ larger than about $0.2$. However,
they all indicate that the walks are localised, with a mean length and
spatial extent which decrease with increasing $p$. These numerical
results thus complement those of section \ref{analytic}.  In addition
to their relevance to the corresponding quantum localisation problem,
these classical processes are of interest in their own right. In
section
\ref{discussion}, we discuss possibilities for related future work.

\section{Analytic results}
\label{analytic}

\subsection{Estimation of the diffusion constant}
\label{diff}
As a first step in analysing the weak-interaction limit of this model,
we estimate the diffusion constant of a random walker on the Manhattan
lattice, subject to no restrictions on return to the same node or link.
A diffusive path ${\bf r}(t)$ which takes a step of finite length at
each unit time exhibits the behaviour
\begin{equation}
\langle \left[ {\bf r}(t)-{\bf r}(0) \right]^2 \rangle = 4Dt\,,
\end{equation}
where the lattice-dependent quantity $D$ is the diffusion
constant. For convenience, we shall take ${\bf r}(0)={\bf 0}$ in what
follows. The diffusion constant for a given random walk may be
calculated by recasting the position coordinates of points visited on
the path as a sum $r_t$ of complex numbers:
\begin{equation}
\label{threeone}
r_t= z_1+z_2+\ldots +z_t\,,
\end{equation}
where $z_{j+1}=z_j\,e^{i\theta_j}$. For the random walks in which we
are interested, we require additionally $|z_j|=a$ for all $j$, where
$a$ is the lattice bond length. The distribution of $\theta_j$ depends
on the details of the lattice; the $\theta_j$ are assumed to be
uncorrelated, so that each step of the random process is independent
of all previous history. For the Manhattan lattice, the distribution
for the $\theta_j$ is somewhat complicated because the allowed turning
directions at a given node are dependent upon the position of the node
on the lattice. For turning probability $p$, the mean free path of a
particle moving randomly on the Manhattan lattice $\tau_{\rm mfp}\sim
1/p$; hence for small $p$, $\tau_{\rm mfp}\gg 1$. We therefore assume that
the particle is unable, on the length scale of the mean free path, to
tell whether it will next change direction at a left-turn or
right-turn node. Thus the options of turning left or right are
assigned equal probability in this approximation. The corresponding
distribution for $\theta_j$ is
\begin{equation}
\theta_j=
\begin{cases}
 \quad 0& \text{ with probability }1-p\\ -\pi/2& \text{ with
 probability }p/2\\ +\pi/2& \text{ with probability }p/2\,,\\
\end{cases}
\end{equation}
with angles measured in an anti-clockwise direction.  Thus $\langle
e^{i\theta_j}\rangle = 1-p + e^{i\pi /2}\,p/2+e^{-i\pi /2}\,p/2=1-p$
for all $j$, and $\langle e^{i(\theta_1+\ldots +\theta_r)}\rangle =
\langle e^{i\theta_1}\rangle^r = (1-p)^r.$ Now from (\ref{threeone}),
we have
\begin{equation}
r_t=z_1\,(1+e^{i\theta_1}+\ldots +e^{i(\theta_1+\ldots
+\theta_{t-1})})\,.
\end{equation}
Hence, since $\langle | {\bf r}(t)|^2 \rangle \equiv \langle r_tr_t^*
\rangle$,
\begin{equation}
\langle | {\bf r}(t)|^2 \rangle =  |z_1|^2
\left((2-p)p^{-1}t-2(1-p)p^{-2}
\left(1-(1-p)^t\right)\right).
\end{equation}
At large $t$, $(1-p)^t\rightarrow 0$, yielding
\begin{equation}
\langle | {\bf r}(t)|^2 \rangle \approx |z_1|^2
\left((2-p)p^{-1}\,t-2\,(1-p)\,p^{-2}\right)\,.
\end{equation}
Thus for the Manhattan lattice,
\begin{equation}
D\approx (2-p)\,p^{-1}\,.
\end{equation}
We note that at $p=0$, the diffusion constant becomes infinite; this
characterises superdiffusive behaviour. In this particular example of
superdiffusion, we also find that the mean square displacement
$\langle \left[ {\bf r}(t)-{\bf r}(0) \right]^2 \rangle$ diverges
faster than linearly with time. In fact, it is easy to see that
$\langle \left[ {\bf r}(t)-{\bf r}(0) \right]^2 \rangle = a^2\,t^2$
since for $p=0$ the random walk proceeds in a straight line. We
therefore see a transition to a different kind of behaviour exactly at
$p=0$.

\subsection{A field theoretic approach}
\label{peliti_theory}
\subsubsection{Peliti's field theory for true self-avoiding walks}
The starting point for our renormalisation group approach is the field
theoretic description proposed by Peliti \cite{amit,obukhov,peliti}
for the class of true self-avoiding walks to which our bond-avoiding
paths on the Manhattan lattice belong. The theory is composed of two
scalar fields $\tilde{\psi}({\bf r},t)$ and $\psi({\bf r},t)$, which
respectively create and destroy walks. The Hamiltonian ${\cal H}$ is
given by ${\cal H}={\cal H}_0+{\cal H}_I$, where ${\cal H}_0$ is the
`free' Hamiltonian
\begin{equation}
\label{h0}
{\cal H}_0=\int_0^\infty {\rm d}t\int {\rm d}^d{\bf r}
\tilde{\psi}({\bf r},t)\, \left[ -\frac{\partial}{\partial
t}+D\,\nabla^2\right]\,\psi({\bf r},t)
\end{equation}
and ${\cal H}_I$ is an interaction Hamiltonian whose precise form
depends on the nature of the process under consideration. $D$
represents the bare diffusion constant, as computed above. From this
class of Hamiltonians arise well-defined Feynman rules for the
theory. The correlation function $G({\bf r},t)=\langle
\tilde{\psi}(0,0) \,\psi({\bf r},t)\rangle$ yields the probability
density for a walk beginning at the origin to be found at position
${\bf r}$ after time $t$. After applying a Fourier transform to the
spatial components of $G({\bf r},t)$ and a Laplace transform to the
temporal component, we obtain the bare propagator
\begin{equation}\tilde{G}({\bf p},\mu)=\tilde{G}_0({\bf p},\mu)\equiv
(\mu + Dp^2)^{-1}\,. 
\end{equation}
The generalised interaction contribution is given by
\begin{equation}
\label{gamma}
\gamma({\bf p},\,{\bf q},\,{\bf p_1})=({\bf p_1}.{\bf q})\,g_1+({\bf
p_1}.({\bf p_1}+{\bf q}))\,g_2+({\bf p_1}.{\bf p})\,g_3\,,
\end{equation}
where ${\bf p}$ is the earliest incoming wavenumber and ${\bf p_1}$ is
the latest outgoing wavenumber. The values of the couplings $g_i$
depend on the nature of the random walk. For example, for Peliti's
true self-avoiding walk, one should take $g_1=g$ and $g_2=g_3=0$; this
corresponds to the interaction Hamiltonian
\begin{equation}
\begin{split}
{\cal H}_I &= -g \int_0^\infty {\rm d}t\, {\rm d}t'\int {\rm d}^d{\bf
r} \left( \psi ({\bf r},t') \nabla \tilde{\psi}({\bf r},t') \right) \\
&\phantom{\qquad \qquad \qquad \qquad \qquad \qquad} \cdot \left(\psi
({\bf r},t) \nabla \tilde{\psi}({\bf r},t)\right)\,.
\end{split}
\end{equation}
Since the coupling constants are dimensionless quantities, the upper
critical dimension of this class of theories is $d_c=2$. The general
theory may be renormalised by dimensional regularisation and minimal
subtraction in two dimensions. Mass renormalisation is
trivial. However, the diffusion constant is renormalised by a factor
$Z$ so that $D=Z\,D_R$, and renormalised coupling constants are given
by $u_i\,\kappa^{\epsilon}$, $i=1,\,2,\,3$, where $\kappa$ is the
renormalisation wavenumber. A calculation to order one loop gives
diffusion constant renormalisation factor
\begin{equation}
\label{Zrenorm}
Z=1+\epsilon^{-1}(u_3-u_1)
\end{equation}
and renormalisation group equations
\begin{align}
&\frac{{\rm d}u_1}{{\rm d}\tau}=-\epsilon u_1+\tfrac{5}{2}u_1^2
+\tfrac{1}{2}u_2^2-2u_1u_2-3u_1u_3\\ &\frac{{\rm d}u_2}{{\rm
d}\tau}=-\epsilon u_2-u_1^2-u_2^2+3u_1u_2+u_1u_3-3u_2u_3\\ &\frac{{\rm
d}u_3}{{\rm d}\tau}=-\epsilon u_3-\tfrac{5}{2}u_3^2+2u_1u_3-u_2u_3
\end{align}
where $\tau=\log (\kappa'/\kappa)$ is the logarithm of the scale
parameter.

\subsubsection{Constructing an interaction Hamiltonian}
In order to apply these general results to the random walks in which
we are interested, we now examine in detail the types of interaction
which might arise for kinetic self-avoiding processes on the Manhattan
lattice. There are two competing interactions involved: the node
interactions are attractive, and clearly the bond interactions in this
type of self-avoiding process must be repulsive. Thus one way to look
at the effect of these interactions is to begin with a free kinetic
walk on the Manhattan lattice and then introduce appropriate 
weights to model self-avoiding characteristics. We impose a repulsive
weight $\exp (-\beta VM(M-1)/2)$ (with $V>0$)
for every bond visited $M$ times and
an attractive weight $\exp (+\beta V'N(N-1)/2)$ (with $V'>0$)
for every node visited
$N$ times, and examine the resultant modification to the free kinetic
walk. To work to first order in $V$ and $V'$, it is sufficient to
consider only cases where a path has visited each node and bond on the
lattice no more than twice.

On average, there is no directional bias to these processes on the
Manhattan lattice; therefore, no drift terms, or equivalently no first
order gradient terms, should appear in the interaction
Hamiltonian. However, higher order gradients may be obtained by
examining the hopping terms generated as the particle moves on the
lattice. These should respect the global rotational invariance of the
kinetic walks. Consider, for example, a move from ${\bf r_1}$ to ${\bf
r_2}$ at time $t_2$. This is effected by the probability-conserving
term $\tilde{\psi}({\bf r_2})\,\psi({\bf r_1})-\tilde{\psi}({\bf
r_1}\,)\psi({\bf r_1})$. Note that the particle density at position
${\bf r}$ and time $t$ is given by $\tilde{\psi}({\bf r},t)\,\psi({\bf
r},t)$, and takes values $0$ or $1$. Now suppose that the midpoints of
the exit bonds at a particular node have position vectors ${\bf r_i}$,
with $i=1,\ldots,4$. Thus to describe a first visit to the node at
time $t_1$ and a subsequent visit at time $t_2$, approaching from
${\bf r_1}$ and leaving via ${\bf r_2}$, we postulate an interaction
term of the form
\begin{equation}
\begin{split}
[\tilde{\psi}({\bf r_2},t_2)\,\psi({\bf r_1},t_2)-&\tilde{\psi}({\bf
r_1},t_2)\,\psi({\bf r_1},t_2)]\\ &\times
\sum_{i=1}^4\,\alpha_i\,\tilde{\psi}({\bf r_i},t_1)\,\psi({\bf
r_i},t_1)\,,
\end{split}
\end{equation}
with the $\alpha_i$ appropriately chosen to give the correct Boltzmann
weights. Suppose that the possible routes through a given node are
${\bf r_1}\rightarrow{\bf r_2}$, ${\bf r_1}\rightarrow{\bf r_3}$,
${\bf r_4}\rightarrow{\bf r_2}$ and ${\bf r_4}\rightarrow{\bf
r_3}$. Then applying weights $V$ and $V'$ as defined above, if the
path passes from ${\bf r_1}$ to ${\bf r_2}$ on both visits to the
node, we obtain
\begin{equation}
\alpha_1+\alpha_2=-2V+V'\,.
\end{equation} 
Considering each of the other possible routes through the node, we
find also
\begin{align}
&\alpha_1+\alpha_3=-V+V'\\ &\alpha_2+\alpha_4=-V+V'\\
&\alpha_3+\alpha_4=+V'
\end{align}
We therefore have a set of simultaneous equations for the
$\alpha_i$. Note that these are invariant under the transformation
$\alpha_{1,4}\rightarrow \alpha_{1,4}+A$,
$\alpha_{2,3}\rightarrow\alpha_{2,3}-A$ for any $A$, so that there is
always some arbitrariness in the solutions. However, if $V'=0$,
i.e. there is no node interaction, then $\alpha_1=\alpha_2=-V$,
$\alpha_3=\alpha_4=0$ is a solution. Similarly, if $V=0$, i.e. there
is no bond interaction, then
$\alpha_1=\alpha_2=\alpha_3=\alpha_4=V'/2$ is a solution. We thus
obtain bond interactions of the form
\begin{equation}
\begin{split}
-V\,(\tilde{\psi}({\bf r_j},t_2)\,&\psi({\bf
r_i},t_2)-\tilde{\psi}({\bf r_i},t_2)\,\psi({\bf r_i},t_2))\\ &\times
(\tilde{\psi}({\bf r_j},t_1)\,\psi({\bf r_i},t_1)+\tilde{\psi}({\bf
r_j},t_1)\,\psi({\bf r_j},t_1))
\end{split}
\end{equation}
and node interactions of the form
\begin{equation}
\begin{split}
\dfrac{1}{2}V'\,(\tilde{\psi}({\bf r_j},t_2)\,\psi({\bf r_i},t_2)-&
\tilde{\psi}({\bf r_i},t_2)\,\psi({\bf r_i},t_2))\\
&\times \sum_{{\bf r'}\in {\text {node}}}\tilde{\psi}({\bf
r'},t_1)\,\psi({\bf r'},t_1)\,,
\end{split}
\end{equation}
where, on the second visit, the path approaches the node from ${\bf
r_i}$ and leaves via ${\bf r_j}$.

\begin{figure}
\begin{center}
\includegraphics[width=5cm]{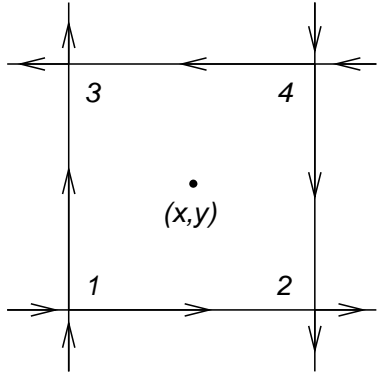}
\caption[A unit cell on the Manhattan lattice]{A unit cell on the
Manhattan lattice with centre at the point ($x$,$y$); the possible
node types are labelled 1 to 4. Each lattice bond has length $2a$.}
\label{fig:cellfig}
\end{center}
\end{figure}
A unit cell on the Manhattan lattice contains four different node
configurations, as illustrated in Fig. \ref{fig:cellfig}. In order to
construct the interaction terms in the Hamiltonian then, we must
compute the contribution for both node and bond interactions from each
of the nodes in the unit cell.  Consider, for example, node 1 in
Fig. \ref{fig:cellfig}, positioned at $(x-a,y-a)$. A possible route
through this node might arrive at the point ${\bf r_j}=(x-2a,y-a)$ at
time $t_2$ and leave via ${\bf r_i}=(x,y-a)$. The second order
gradient expansion for this hopping process is given by
\begin{equation}
\begin{split}
&\tilde{\psi}({\bf r_j},t_2)\psi({\bf r_i},t_2)-\tilde{\psi}({\bf
r_i},t_2)\psi({\bf r_i},t_2)\\
&=\tilde{\psi}(x,y-a,t_2)\psi(x-2a,y,t_2)\\ &\qquad \qquad \qquad
-\tilde{\psi}(x-2a,y,t_2)\psi(x-2a,y,t_2)\\ &=2a\,\frac{\partial
\tilde{\psi}_2}{\partial x}\,\psi_2-2a^2 \frac{\partial ^2
\tilde{\psi}_2}{\partial x^2}\psi_2-2a^2\frac{\partial ^2
\tilde{\psi}_2}{\partial x \partial y}\psi_2\\
&\qquad \qquad \qquad \qquad -4a^2\,\frac{\partial
\tilde{\psi}_2}{\partial x}\frac{\partial \psi_2}{\partial x}
-2a^2\frac{\partial \tilde{\psi}_2}{\partial x}\frac{\partial
\psi_2}{\partial y}\,.\\
\end{split}
\end{equation}
In the last line above, the subscript $2$ indicates time $t_2$ and the
position argument $(x,y)$ of the $\psi$, $\tilde{\psi}$ has been
omitted for clarity.  Combining the contributions for all the possible
routes through each of the four nodes belonging to the unit cell, we
obtain the node interaction term
\begin{equation}
\begin{split}
\label{node1}
-16\dfrac{V'}{2}a^2(&3(\nabla \tilde{\psi}({\bf r},t_2).\nabla
\psi({\bf r},t_2))( \tilde{\psi}({\bf r},t_1) \psi({\bf r},t_1))\\
&+4\nabla .(\sigma_z(\nabla \tilde{\psi}({\bf r},t_2))\psi({\bf
r},t_2) \tilde{\psi}({\bf r},t_1)\psi({\bf r},t_1)))\\
\end{split}
\end{equation}
and bond interaction term
\begin{equation}
\label{bond1}
\begin{split}
8Va^2(&3(\nabla \tilde{\psi}({\bf r},t_2).\nabla \psi({\bf r},t_2))(
\tilde{\psi}({\bf r},t_1) \psi({\bf r},t_1))\\ &+4\nabla
.(\sigma_z(\nabla
\tilde{\psi}({\bf r},t_2))\psi({\bf r},t_2) \tilde{\psi}({\bf r},t_1)
\psi({\bf r},t_1)))\,,
\end{split}
\end{equation}
where $\sigma_z$ is the Pauli matrix with entries
$\sigma_{ij}=1-\delta_{ij}$.  The interaction Hamiltonian may now be
obtained by integrating over all position space and over $t_1$ and
$t_2$ with $0<t_1<t_2$. Note that the total divergences in
(\ref{node1}) and (\ref{bond1}) simply yield a boundary term on
integrating with respect to ${\bf r}$. We therefore have
\begin{equation}
\begin{split}
\label{H_int}
{\cal H}_I=-24(V'-V)\int{\rm d}^d{\bf r}&\int_0^{\infty}{\rm
d}t_2\int_0^{t_2}{\rm d}t_1(\nabla\tilde{\psi}({\bf r},t_2)\\ &\cdot
\nabla\psi({\bf r},t_2))(\tilde{\psi}({\bf r},t_1)\psi({\bf
r},t_1))\,.
\end{split}
\end{equation}
We see immediately, by comparison with the free Hamiltonian ${\cal
H}_0$ in (\ref{h0}), that the effect of this interaction is only to
modify the diffusion constant. Thus if the trail has already visited a
particular region of the lattice, it will experience a change in
diffusion speed when it revisits this region. The nature of the
modification depends solely upon the relative values of the bond and
node interaction strengths, through the factor $V'-V$. The interaction
derived above leads to a coupling proportional to ${\bf p_1}.({\bf
p_1}+{\bf q})$. Therefore, referring back to the generalised
interaction term displayed in (\ref{gamma}), we see that in order to
obtain the correct coupling for the kinetic self-avoiding trail on the
Manhattan lattice, we must choose initially $g_2=O(V,V')$, and
$g_1=g_3=O(V^2,{V'}^2,VV')$. 
This is in contrast to the kinetic self-avoiding trail
considered by Peliti, for which $g_1>0$ and $g_2=g_3=0$ at $t=0$. We
see from (\ref{H_int}) that, in the Manhattan case, $g_2$ is proportional
to $V'-V$, to first order.

\subsubsection{Renormalisation group flows}
If $g_3=0$ initially, then to first order, $u_3=0$ always. 
To simplify matters, we assume that this remains the case: this does
not affect our qualitative conclusions.
If we also
choose $u_1=0$ initially, then from Eq. \ref{Zrenorm} we find that the
diffusion constant renormalisation factor $Z=1$ at this point. This
indicates that 
there is no singular contribution to the diffusion constant
from the renormalisation at weak coupling,
and therefore the dynamic exponent $z=2$, 
with the possible appearance of 
corrections from the strong coupling behaviour; that is, 
time $t$ scales as the square of the spatial
extension $r=|{\bf r}(t)|$. From the analysis presented earlier in
this section, in $d=2$ dimensions the renormalisation group equations
for $u_1$ and $u_2$ in the $u_3=0$ plane are given by
\begin{align}
\label{flow1}&\frac{{\rm d}u_1}{{\rm d}\tau}=\tfrac{5}{2}u_1^2 +
\tfrac{1}{2}u_2^2-2u_1u_2\\
\label{flow2}&\frac{{\rm d}u_2}{{\rm d}\tau}=-u_1^2-u_2^2+3u_1u_2
\end{align}
Hence the only fixed point for $\epsilon=0$ is the trivial fixed point
$u_1=u_2=0$. This corresponds to the case $p=0$, for which the trails
are necessarily non-interacting. The renormalisation group flow
equations may be integrated to yield the general solution
\begin{equation}
u_1-u_2=A(u_1+u_2)^{5/3}\,(2u_1-u_2)^{1/3}\,,
\end{equation}
where if initially $u_1=0$ and $u_2=u_2^0$, then $A=1/u_2^0$. Much may
be learned by examining these flow trajectories.
\begin{figure}
\begin{center}
\includegraphics[width=10cm]{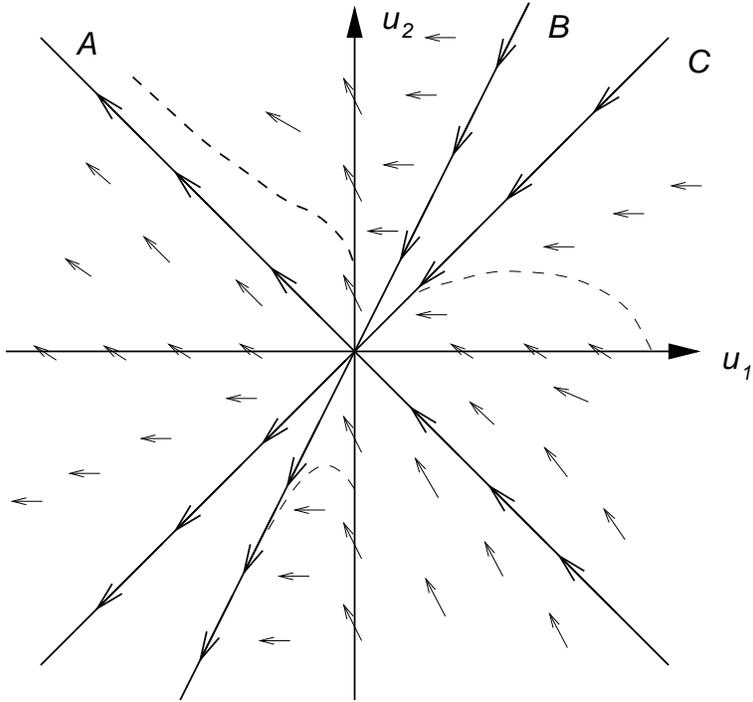}
\caption[Schematic renormalisation group flow diagram in the plane
$u_3=0$]{Schematic renormalisation group flow diagram in the plane $u_3=0$. 
The arrows indicate the direction of decreasing $\tau$,
i.e. increasing $\kappa$. The three attractors are labelled (A)
$u_2=-u_1$, (B) $u_2=2u_1$ and (C) $u_2=u_1$. Possible trajectories
are illustrated schematically by dashed lines.}
\label{flow_diagram}
\end{center}
\end{figure}
Fig. \ref{flow_diagram} illustrates schematically renormalisation
group flow in the plane $u_3=0$. The attractors $u_2=Mu_1$ may be
identified by dividing (\ref{flow1}) by (\ref{flow2}) and solving the
resulting equation for $M$. There are three solutions: $u_2=-u_1$,
$u_2=2u_1$ and $u_2=u_1$.

If asymptotically $u_2=Mu_1$ for some $M$, then in this limit the
renormalisation group equations give $u_1,\,u_2\sim \tau^{-1}$. The
flows in the direction of decreasing $\tau$ in Fig. \ref{flow_diagram}
should show the approach to asymptotic behaviour: see
\cite{obukhov}. 
We observe that a trajectory starting at
$u_1\approx0$ and $u_2^0>0$ flows to the line $u_2=-u_1$, whereas a
trajectory beginning at $u_1\approx0$ and $u_2^0<0$ flows to $u_2=2u_1$. In
both cases, we obtain runaway with a flow to strong coupling. 
These two basins of attraction are divided by the separatrix 
$u_2=u_1$, along which the flows also go to strong coupling. 
Although in all three cases the flows go out of the region of validity
of the one-loop approximation, nevertheless their topology is
consistent with what we might expect for the phase diagram of trails
with attractive on-site interaction, depending on the relative strengths of
the bond repulsion $V$ and the node attraction $V'$. Recall that, to first
order,  $u_2^0\propto V-V'$. For $V\gg V'$ we would expect the walks to
be in the universality class of trails, which is the same as that of
ordinary self-avoiding walks\cite{guttmann85b,guttmann85c}. 
We may therefore interpret flows along the attractor $u_2=-u_1$ as 
going towards
a strong-coupling fixed point which corresponds to the universality
class of SAWs, even though this point is beyond the range of
applicability of our analysis. 
Similarly, for $V'\gg V$ we might expect the walks to be collapsed, so
that we may interpret the attractor $u_2=2u_1$ as flowing towards 
a (trivial) strong-coupling fixed point representing compact objects.
Within this interpretation, then, flows along the separatrix go towards
a strong-coupling fixed point corresponding to the collapse transition,
that is the theta-point. (We note that for this simple phase diagram
we need to assume that the initial value of $u_1$, computed to second
order in $V$ and $V'$, is slightly negative,
otherwise there will be an intermediate phase in which the flows go to
weak coupling, as for the Peliti model.)

Although this analysis is based on a weak-coupling expansion, the
topology of the consequent phase diagram should extend to strong
coupling, unless there are further unexpected phase boundaries.
It does not, however, tell us the form of the phase boundary for larger
values of $V$ and $V'$, and in particular in which phase the original lattice
model lies. Thus we turn to numerical simulations in the next Section.

Although the flows in (\ref{flow1},\ref{flow2}) always go to strong
coupling in our case, if we assume that they go to a fixed point
describing a phase with a finite length scale $\xi$ (for example
the radius of gyration in the collapsed phase), we may still use
them to predict the form of the dependence of $\xi$ on the initial,
weak-coupling, parameters. This is because the largest contribution
comes from the neighbourhood of the weak-coupling fixed point.
In general, any such length scale satisfies
\begin{equation}
\xi(u_2^0)=\xi_0 \exp \left( \int_0^{\tau^*}{\rm d}\tau \right)\,,
\end{equation}
where $\xi_0$ is of the order of the microscopic cut-off, $\tau$
is the RG `time', and  $\tau^*$ is chosen so that 
$\xi\big(u_2(\tau^*)\big)=O(1)$.
The integral here may be rewritten as
\begin{equation}
\int \frac{{\rm d}\tau}{{\rm d}u_2}\,{\rm d}u_2 =
	 \int_{u_2^0}^{O(1)} \frac{{\rm d}u_2}{-u_1^2-u_2^2+3u_1u_2}\,.
\end{equation}
Since initially $u_1=O(u_2^2)$, the integral is $O(1/u_2^0)$ as
$u_2^0\to0$, so that 
$\xi \sim \exp (1/u_2^0)$. 
This result would be academic were it not for the fact that the
diffusion constant behaves as $p^{-1}$ as $p\to0$, and this places
us squarely in the weak-coupling regime. 
Denoting the dimensionality of a quantity $A$ by
$[A]$, we find from (\ref{h0}) that $[\tilde{\psi}({\bf r},t)\psi({\bf
r},t)]=k^d$ and $[D]=\mu k^{-2}$, where $\mu$ is the
Laplace-transformed temporal variable and $k$ represents
momentum. From (\ref{H_int}), we see that $[g_2]\,k^{2d}k^2=\mu^2k^d$,
i.e. $[g_2]=\mu^2k^{-d-2}$. Hence $g_2\sim D^2k^{\epsilon}$. Therefore
the renormalised, dimensionless coupling $u_2\sim g_2\, D_R^{-2}
\kappa^{-\epsilon}$. Now since, from section \ref{diff}, the bare
diffusion constant $D
\sim p^{-1}$, we have $u_2^0 \sim p^2$. This leads to the leading
order prediction $\xi \sim e^{{\rm const.}/p^2}$: 
the correlation length increases
exponentially with decreasing $p^2$.
This is entirely consistent with the only critical point of the lattice
model being at $p=0$.

\section{Simulations}
\label{simulations}

\subsection{Preliminaries}
\label{preliminaries}
In this section, we use Monte Carlo simulations to study various
characteristics of kinetic self-avoiding processes on the Manhattan
lattice. The lattice bond length is chosen to be two units, so that
$a=1$ in the analysis above, and the lattice size is effectively
infinite. The walker moves from one bond centre to an allowed
neighbouring bond centre at each step, subject to the lattice bond
directions and the condition that the path be self-avoiding; the
number of steps taken is denoted by $N$. Thus a path consisting of one
step, for which $N=1$, connects two adjacent bond centres and has
end-to-end length two units. Each simulation consists of a batch of
$10^5$ walks, which are grown individually. When a walk reaches the
specified maximum length or closes, that is, returns to its initial
coordinates, it is terminated and a new walk is started; in this way,
the walks sampled at a given length $N$ should be statistically
independent. Data points for different values of $N$ are generated
from separate simulation runs. Errors on each data point are
calculated as the variance of the quantity in question. Where curves
have been fitted to the data, the fitting procedure employed an
implementation of the nonlinear least-squares (NLLS)
Marquardt-Levenberg algorithm.

\subsection{Survival probability}
\label{survival}
\begin{figure}
\begin{center}
\rotatebox{-90}{\includegraphics[width=10cm]{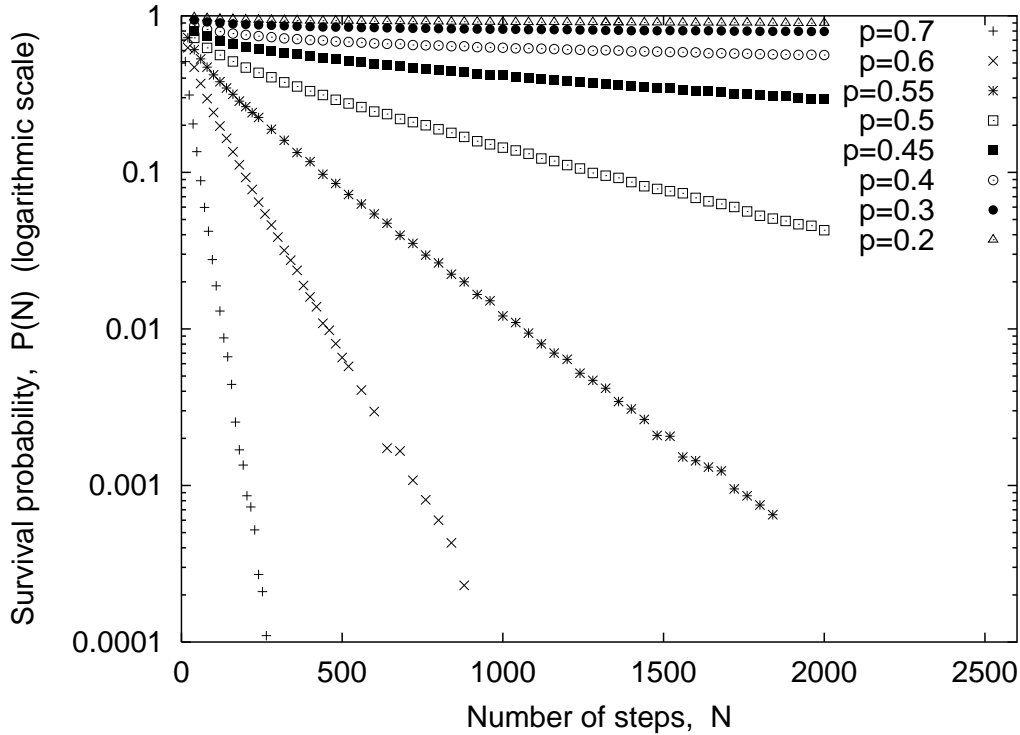}}
\caption[Survival probability as a function of path length]{Survival
probability as a function  of path length, plotted for various
$p$. Maximal values for $N$ are:  $p=0.7$, $N_{max}=264$; $p=0.6$,
$N_{max}=880$; $p=0.55$, $N_{max}=1840$;  $p \leq 0.5$,
$N_{max}=2000$. The vertical axis carries  a logarithmic scale.}
\label{survival_graph}
\end{center}
\end{figure}

Survival probability $P(N)$ is defined as the proportion of simulated
paths which have not yet closed after the $N$th step. This is expected
to be a $p$-dependent quantity: at $p=1$, $P(N)=0$ for $N\geq 4$, and
at $p=0$, $P(N)=1$ for all $N$ as the trails simply proceed straight
along a given direction. The data plotted in Fig. \ref{survival_graph}
suggest that P(N) decays exponentially with increasing $N$, on a
$p$-dependent length scale.
\begin{figure}
\begin{center}
\rotatebox{-90}{\includegraphics[width=10cm]{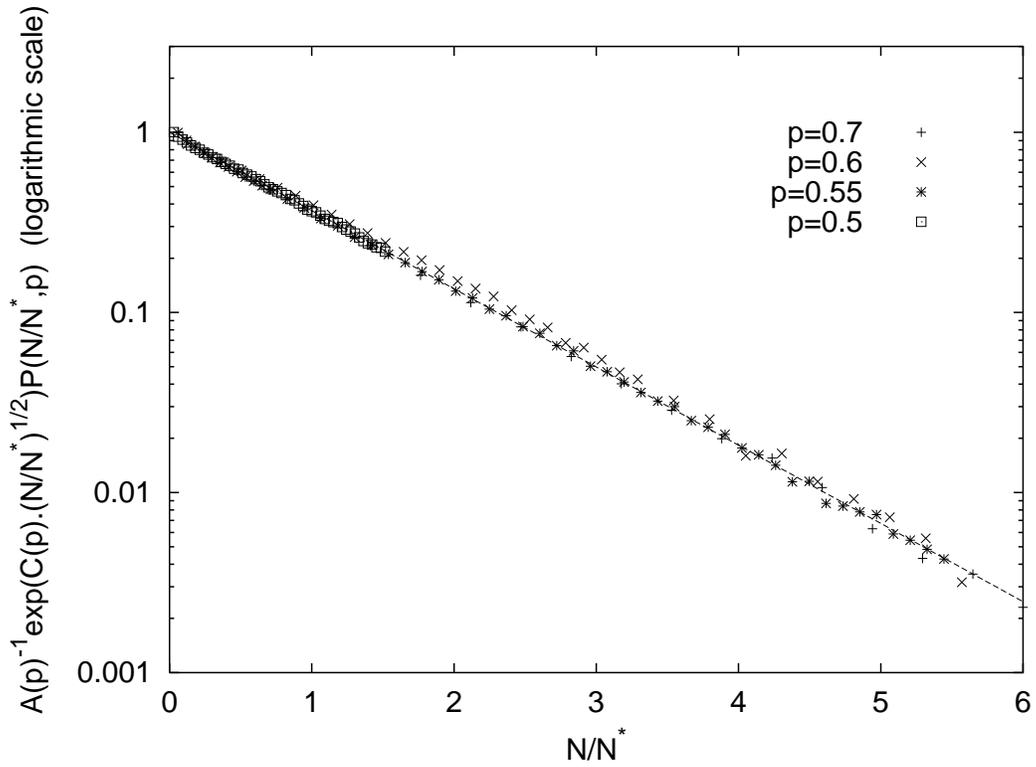}}
\caption[Analysis of survival probability]{Rescaled data for survival
probability, taking the form
$A(p)^{-1}\,P(N/N^*,p)\,\exp(C(p)\sqrt{N/N^*(p)})$, are plotted
against the rescaled path length $N/N^*$ for selected values of
$p$. The dashed line shows $f(N/N^*)=\exp(-N/N^*)$.}
\label{survival_collapse}
\end{center}
\end{figure}
This corresponds to localisation for all non-zero $p$. Since all paths
are extended at $p=0$, this provides evidence in support of the
existence of a trivial critical point, as discussed earlier. In fact,
we find that the simulated values for survival probability are
consistent with a scaling form given by
\begin{equation}
\label{P_scale}
P(N/N^*,p) \sim e^{-N/N^*(p)}\,e^{-C(p)\,\sqrt{N/N^*(p)}}\,,
\end{equation}
where $N^*(p)$ may be interpreted as the typical number of steps taken
by a path before closure. Estimated values for $C(p)$ and $N^*(p)$,
obtained by fitting the scaling form of (\ref{P_scale}) to the
simulated data, are presented in Table \ref{table1} for certain values
of $p$. Note that the typical length $N^*$ increases as $p$ decreases;
for this reason, it was found to be extremely difficult to obtain
reliable estimates for $N^*(p)$ and $C(p)$ for $p<0.5$. Indeed, at
$p<0.5$ it is not possible to give reliable estimates for $N^*$ since
the estimated values are larger than the maximal value of $N$ used in
the simulations. However, this is to be expected since as
$p\rightarrow 0$, $N^*$ should tend to infinity. Fig.
\ref{survival_collapse} shows rescaled data 
$A(p)^{-1}\,P(N/N^*,p)\,\exp(C(p)\sqrt{N/N^*(p})$ plotted against
$N/N^*(p)$, where $A(p)$ is the normalisation factor required for
(\ref{P_scale}). The data collapse on to the line
$f(N/N^*)=\exp(-N/N^*)$, also plotted in Fig.
\ref{survival_collapse}, showing excellent agreement with the
postulated scaling equation. Notice that since the typical length scale
for closure of walks is much longer for smaller values of $p$, as
predicted in the final part of section \ref{peliti_theory}, data for
these simulations is plotted over a smaller range of $N/N^*$. The
scaling behaviour of (\ref{P_scale}) may be justified by the following
argument, which is motivated partially by evidence presented in
section
\ref{internal_energy} for scaling of internal energy.
\begin{table}
\begin{center}
\begin{tabular}{|l|l|l|}
\hline
$p$ & $N^*(p)$ & $C(p)$\\
\hline
0.7 & 34.0 & 0.346\\ 0.6 & 158 & 1.15\\ 0.55 & 338 & 0.748\\ 0.5 &
1320 & 1.23\\
\hline
\end{tabular}
\caption[Values for coefficient $C(p)$ and typical path length
$N^*(p)$]{Values for coefficient $C(p)$ and typical path length
$N^*(p)$, tabulated for various $p$. Accuracy is to 3SF.}
\label{table1}
\end{center}
\end{table}

The kinetic self-avoiding processes in which we are interested are non
equilibrium problems. However, it is possible to use statistical
mechanics to examine their structure by exploiting an equivalence with
certain equilibrium objects. To demonstrate this equivalence, we
employ arguments similar to those used by Bradley for kinetic growth
walks on the Manhattan lattice \cite{bradley1,bradley2} and by
Owczarek and Prellberg for trails on hypercubic lattices
\cite{trails1,trails2}. Consider an equilibrium model whose
configurations are kinetic self-avoiding trails laid down on the
Manhattan lattice. For the moment, we look only at the case
$p=1/2$. The temperature is fixed at a particular value, as is usual
with Monte-Carlo simulations. Now suppose that each doubly occupied
node on a trail is assigned a weighting $w$.  If $w=1$, then this
weighting will simply yield ordinary trails, on which a node may be
visited any number of times. However, if $w$ is large then the path
should be a highly compact object. For our kinetic self-avoiding
trails, the appropriate weighting is $w=2$: when the path arrives at a
node for the first time, it may choose either of two possible exits,
with probability $p=1/2$ of making each choice, but on revisiting the
node, the path must leave by the one unoccupied exit with probability
1. Let $\phi(N)$ be an open path configuration with length $N$ steps,
on which $n(\phi)$ nodes are doubly occupied. The probability of
obtaining this configuration is
$p_{\phi(N)}=(1/2)^{N-n(\phi)}$. However, as we have seen, there is
also the possibility of generating a path which is a closed loop; the
probability $P(N)$ of obtaining an open configuration at length $N$ is
less than 1. In fact,
\begin{equation}
\begin{split}
P(N)&=\sum_{\phi(N)}\, p_{\phi(N)}=
2^{-N}\,\sum_{\phi(N)}\,2^{n(\phi)}\\ &=2^{-N}\,{\cal Z}_N(2)\,,\\
\end{split}
\label{open_P}
\end{equation}
where ${\cal Z}_N(2)$ is the partition function for the equilibrium
model, with the Boltzmann weight $w=2$. Thus the set of kinetic
self-avoiding growth trails at $p=1/2$ is a sample from the Boltzmann
distribution of equilibrium trails with weighting $w=2$. Further, the
mean number of doubly occupied nodes is given by $w\,{\rm d}F/{\rm
d}w$ evaluated at $w=2$, where $F$ is the free energy of the
equilibrium model. This relationship is particularly useful since the
mean number of doubly occupied nodes may readily be computed from
simulations.

For a free random walk, ${\cal Z}_N=4^N$, where $N$ is the number of
steps taken, since the walker may leave a given node via any of the
four associated bonds. Thus, in this case, $F\sim N\log 4$. Note that
since the bulk fractal dimension of a walk model is given by
$d_f=1/\nu$, where the radius of gyration scales with exponent $\nu$,
here $d_f=2$. However, the free random walk does not have a well
defined surface. In contrast, there are other types of walk for which
a surface may be identified. For collapsed polymers, for example,
which may be modelled by certain varieties of compact walk, a term
appears in the partition function ${\cal Z}$, and hence also in the
free energy $F$ which scales according to the surface area of the
object \cite{opb}. For an object embedded in $d$ dimensions, the
average extension in any given spatial direction is given by $R\sim
N^{1/d}$, since $\nu=1/d$. Therefore the surface area must scale as
$R^{d-1}\sim N^{(d-1)/d}$. Thus in two dimensions, the surface area of
a compact object scales like $N^{1/2}$. This leads to an expression of
the free energy of the form
\begin{equation}
\label{free}
F\sim NF_b + N^{1/2}F_s+\ldots\,,
\end{equation}
where the $\ldots$ indicates smaller correction terms, and $F_b$ and
$F_s$ are respectively bulk and surface contributions to the free
energy.  Hence the partition function must scale as
\begin{equation}
\label{Z_scale}
{\cal Z}_N(2) \sim \exp(NF_b + N^{1/2}F_s+\ldots)\,.
\end{equation}
From (\ref{open_P}), we may therefore deduce that if the kinetic
self-avoiding trails behave as compact walks, the probability of
finding an open configuration after $N$ steps should scale as
\begin{equation}
\label{P_scale2}
P(N) \sim \exp(N\log 2 +NF_b + N^{1/2}F_s+\ldots)\,.
\end{equation}
This is consistent with the scaling form of (\ref{P_scale}) which was
fitted to the simulated data, suggesting that the kinetic
self-avoiding trails are compact objects. It follows from our
postulation of a trivial critical point that the scaling form
(\ref{P_scale2}) put forward for the case $p=1/2$ should hold for all
non-zero $p$.

We note that our results indicate that kinetic growth trails on the
Manhattan lattice are mapped to a temperature inside the collapsed
phase of the equilibrium model. Previous kinetic growth models of
walks and trails \cite{bradley1,bradley2,trails1,trails2} in two and
higher dimensions have all mapped to temperatures in the extended
phase or $\theta$-point of the corresponding equilibrium model. This
makes our kinetic growth model of interest from the point of view of
polymer models.

\subsection{Radius of gyration and end-to-end distance}
\label{rg_and_re}
\begin{figure}
\begin{center}
\rotatebox{-90}{\includegraphics[width=10cm]{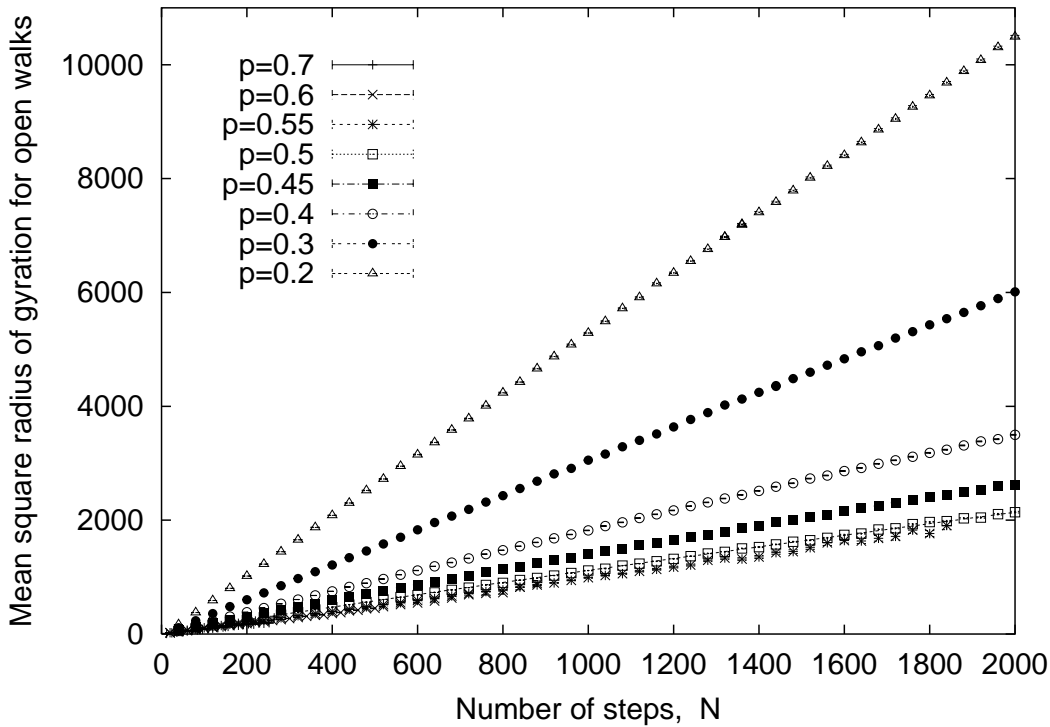}}
\caption[Scaling of mean square radius of gyration for open
trails]{Scaling of mean square radius  of gyration $\langle R_G^2
\rangle$ for open trails sampled after $N$  steps. Data are plotted
for $p$ between 0.2 and 0.7. The line fitted to the data for $p=0.5$
has equation $\langle R_G^2 \rangle \propto N^{0.94}$, where the
exponent is quoted to two significant figures.}
\label{rgopen}
\end{center}
\end{figure}

In this section we consider scaling of the spatial extension of the
kinetic self-avoiding trails. We begin by examining the mean square
radius of gyration, $\langle R_G^2 \rangle$, which is expected to
scale as $N^{2\nu}$, for some $\nu$. For a non-interacting random
flight, $\nu=1/2$: $\langle R_G^2 \rangle = Na^2\left(
1-N^{-2}\right)/6$, where $a$ is the (constant) step length
\cite{hughes1}. It is also known that $\nu=1/d$ for collapsed polymers
in $d$ dimensions.  Fig. \ref{rgopen} shows simulated data for trails
which are still open after $N$ steps.
On fitting a scaling form $\langle R_G^2 \rangle = A(p)\,N^{2\nu(p)}$
to each data set, we find that the estimated values for the exponent
$\nu$ lie in the range $0.47 \lessapprox \nu \lessapprox 0.52$. Of
course since the trails can never visit any bonds twice one can easily
show that $\nu \geq 1/2$. We have also studied the mean square radius
of gyration for walks which close at the $N$th step; the simulated
data support a scaling form $\langle R_G^2 \rangle \sim N$, although
the statistics obtained from these walks are not high.  For example,
for closed trails at $p=0.5$, we have approximately $\langle R_G^2
\rangle \propto N^{1.05}$.  This evidence strongly suggests that for
both open and closed trails, $\langle R_G^2 \rangle \sim N^{2\nu}$
with $\nu=1/2$. This implies that the trails should be classified
either as non-interacting random flights (polymer chains) or as
compact walks (polymers in the collapsed phase) since both these have
$\nu=1/2$ in two dimensions. However, evidence from the previous
section suggests that the compact hypothesis is the most consistent
one.

\begin{figure}
\begin{center}
\rotatebox{-90}{\includegraphics[width=10cm]{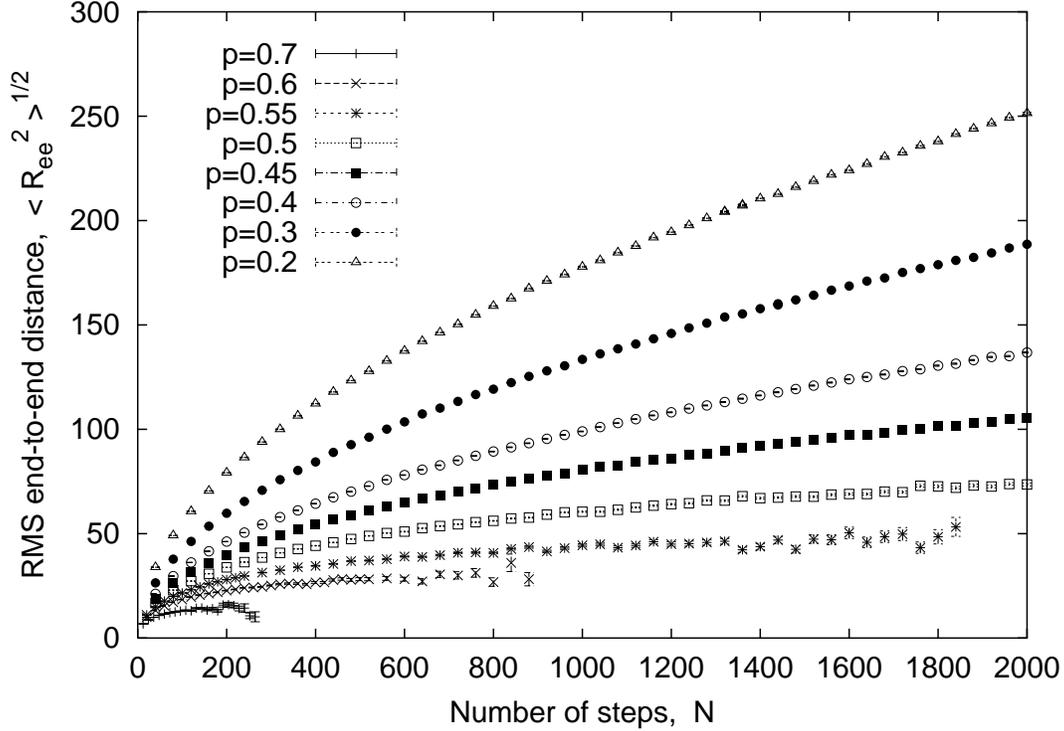}}
\caption[Scaling of RMS end-to-end distance]{Scaling of RMS end-to-end
distance with number of steps, for open trails, at various $p$. A line 
fitted to the data for $p=0.5$ has equation $\langle R_e^2 \rangle
\approx 0.104N + 135N^{1/2} - 728$, showing linear scaling with
corrections $O(N^{1/2})$.}
\label{ree}
\end{center}
\end{figure}

We next consider the scaling of end-to-end distance $R_e$ for the
trails. For the free random flight model, $\langle R_e^2 \rangle =
Na^2$, where, as before, the step length is given by $a$. Thus for
large $N$, $\langle R_G^2 \rangle \sim \langle R_e^2 \rangle /6$ in
this model. For compact walks, however, there should be corrections to
this scaling form which result in suppression of the end-to-end
distance relative to the non-interacting case. Fig. \ref{ree} shows
simulated data for the root mean square end-to-end distance
$\sqrt{\langle R_e^2(N) \rangle}$ for trails at various values of
$p$. Since this quantity does not make sense for paths which have
closed, the average is taken only over those paths which are still
open after the $N$th step. We find that the data are consistent with a
scaling form $\sqrt{\langle R_e^2(N) \rangle} \sim
N^{2\nu}\,(1+BN^{-1/2}+CN^{-1})$, where $\nu=1/2$ as above. That is,
we observe linear scaling with corrections of order $N^{1/2}$ and
smaller.
The presence of the correction terms is consistent with behaviour
expected from compact random walks, in contrast to the exact linear
scaling which would obtain if the walks were free random flights. From
the result $\xi \sim e^{1/p^2}$ in section \ref{peliti_theory}, we
expect to be able to conclude that $R_e^2/N \sim e^{1/p^2}$. However,
this is difficult to verify.

\subsection{Internal energy}
\label{internal_energy}
In the equilibrium model described in section \ref{survival}, we may
define the average internal energy $U$ of a configuration of $N$ steps
by
\begin{equation}
\begin{split}
\langle U \rangle &= 
-\frac{\partial}{\partial
\beta}\,\log{\cal Z}_N(w)\\
&\sim -\frac{\partial}{\partial \beta}\,F_N(w)
\end{split}
\end{equation}
since, up to constants, $F_N\sim\log{\cal Z}_N$.  Since $F_b$ and
$F_s$ in (\ref{free}) are expected to be temperature dependent, taking
the derivative of $F$ with respect to temperature gives
\begin{equation}
\label{uscale}
u\sim u_b+ u_s N^{-1/2} +\ldots
\end{equation}
for the internal energy per step, $U/N$. Here, the corrections, which
have carried through from the expression for free energy, are expected
to be of the order of $N^{-1}$. The subscripts $b$ and $s$ denote bulk
and surface terms, as previously.

We now compare these results with simulations conducted on the
Manhattan lattice.
\begin{figure}
\begin{center}
\rotatebox{-90}{\includegraphics[width=10cm]{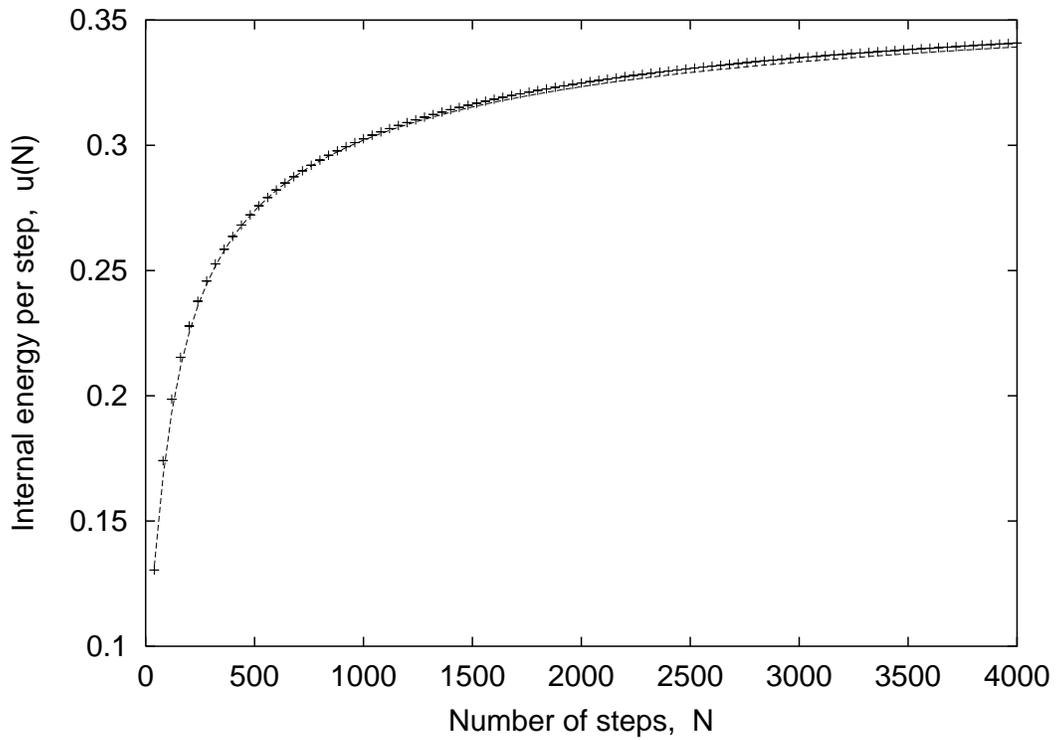}}
\caption[Scaling of internal energy per step, $u$]{Scaling of internal
energy per step, $u$, with path length, for trails with $p=0.5$ open
at length $N$, up to maximum length $N=4000$. The line fitted has
equation $u(N)=0.38-2.69N^{-1/2}+7.09N^{-1}$; the coefficients are
accurate to two decimal places.}
\label{uint}
\end{center}
\end{figure}
Fig. \ref{uint} shows data for internal energy per step, $u(N)$,
generated from populations of open trails of length $N$ steps at
$p=1/2$. Application of a scaling form of the type given in
(\ref{uscale}) indicates that these kinetic trails exhibit behaviour
consistent with the existence of a well-defined one-dimensional
surface. We find the asymptotic behaviour $u\sim u_b\approx 0.38$,
with $u_s \approx 2.7$, and as predicted, the corrections to the bulk
and surface terms are of the order of $N^{-1}$. Since the bulk fractal
dimension for these objects is $1/\nu=2$, we conclude that kinetic
self-avoiding trails on the Manhattan lattice are compact, rather than
extended, with a form similar to that of a liquid droplet. This
conclusion is supported by the scaling forms already presented for
survival probability, end-to-end distance and radius of
gyration. Although the above analysis was carried out for the
particular case $p=1/2$, the same conclusions should hold for all
trails away from the fixed point $p=0$, since all trails generated at
$p>0$ belong to the same phase.

\section{Conclusions and Discussion}
\label{discussion}
We have presented a model for quantum localisation in class C, on
the Manhattan lattice. The general results of \cite{bcc}
show that this may be mapped to a classical random walk problem,
of trails with an attractive on-site interaction. The model depends
on a single parameter $p$. A weak-coupling
field-theoretic renormalisation group analysis appropriate
for small $p$, numerical results for moderate values of $p$, and a
rigorous argument for $p>\frac12$, all lead to the conclusion that
for all $p>0$ the walks almost always
close after a finite number of steps, and correspond to compact objects
with a typical linear size which is a monotonically decreasing function
of $p$. As $p\to0$, this size diverges like $\exp({\rm const.}/p^2)$.

This picture leads to the conclusion that the quantum model on the
Manhattan lattice is always in the localised phase, in contrast to
the case of the L-lattice. This might have been expected, given the
absence of edge states in the former case. We note that the
weak-coupling RG equations for the classical walks have a critical
dimension of two, as is the case for Anderson localisation. It would be
interesting to compare the RG beta-function along the attractor
$u_2=2u_1$ with that obtained using sigma model methods \cite{alt-zirn}.
This would involve relating our coupling $u_2$ to the 
conductance, which is the physical coupling of the sigma model approach.

The arguments above directly apply to the dynamic model of kinetic
growth trails on the Manhattan lattice. However, an equivalence exists
to a particular temperature of a (static) equilibrium statistical
mechanical model of polymers on that lattice, namely self-interacting
trails.  Hence, the results allow us to conclude that this temperature
lies in the collapsed phase of the equilibrium model. This is the
first time a kinetic growth walk/trail model has mapped into such a
phase.

There are several possibilities for extension of this work. The
mapping established in \cite{bcc} requires each node to have
coordination number 4, with two incoming bonds and two outgoing
bonds. This means that every node must either be of the type found on
the L lattice, where a random walker is required to make a turn either
to the left or to the right, or of the Manhattan type, where the
walker may either make a turn or continue straight on. In this paper,
we have considered only graphs consisting of nodes of one type;
however one could also consider graphs consisting of an arrangement of
nodes of both types. This work would be interesting in terms of both
the quantum localisation behaviour and the nature of the corresponding
classical walks. In \cite{bcc}, Sp(4) bilayer systems were
discussed, consisting of two coupled lattice layers each with Sp(2)
symmetry. Work on Sp(4) models of this type is in progress
\cite{bcc2}. In this vein, one could examine mixed bilayer systems, in
which the upper and lower layers have different lattice
structures. For example, the localisation behaviour of a model with
one L lattice layer and one Manhattan lattice layer might be expected
to differ from that of a system consisting of two L lattice
layers. The results presented in this paper and in \cite{bcc}
relate to models based in two dimensions. However, another possibility
for further work would be to consider models whose underlying graphs
are embedded in three or more dimensions. For example, in three
dimensions, one could study a layered system, such as the
three-dimensional U(1) model constructed by Chalker and Dohmen
\cite{chalker-dohmen}. For systems in three or more dimensions, both
insulating and metallic phases should exist; it is expected that each
type of behaviour should hold over a range of values of $p$, with a
transition from one to the other at some critical value $p=p_c$, where
$p_c \in [0,1]$. It would be interesting to discover whether this is
indeed the case.

\section*{Acknowledgements}
We thank John Chalker for valuable discussions. The work was supported
in part by the EPSRC under Grant GR/J78327 and by the Australian
Research Council.

\end{document}